\documentclass[aps,pra,reprint,superscriptaddress]{revtex4-1}

\usepackage{graphics}
\usepackage{graphicx}
\usepackage{subfigure}
\usepackage{amsmath}
\usepackage{verbatim}
\usepackage{soul} 	
\usepackage{color} 			
\newcommand{\addEW}[1]{\textcolor{red}{#1}}
\newcommand{\addAS}[1]{\textcolor{black}{#1}}
\newcommand{\addPDS}[1]{\textcolor{black}{#1}}

\begin{document}


\title{Spontaneous parametric down-conversion induced by optomechanical gradient forces in nanophotonic waveguides}


\author{Mohamed Ashour}
\affiliation{Robert Bosch GmbH, Corporate Research and Advance Engineering, 71272, Renningen, Germany}
\email[]{peter.degenfeld-schonburg@de.bosch.com}
\affiliation{University of Konstanz, Department of Physics, 78457 Konstanz, Germany}

\author{Jan Niklas Caspers}
\affiliation{Robert Bosch GmbH, Corporate Research and Advance Engineering, 71272, Renningen, Germany}

\author{Eva M. Weig}
\affiliation{University of Konstanz, Department of Physics, 78457 Konstanz, Germany}
\affiliation{Technical University of Munich, Department of Electrical and Computer Engineering, 80333 Munich, Germany}
\email[]{eva.weig@tum.de}

\author{Peter Degenfeld-Schonburg}
\affiliation{Robert Bosch GmbH, Corporate Research and Advance Engineering, 71272, Renningen, Germany}


\date{\today}

\begin{abstract}
	Optomechanical gradient forces arise from evanescent fields of guided waves in parallel photonic waveguides. When designed to be of attractive nature, they increase exponentially as the gap between the waveguides decreases. Moreover, the amplitude of the gradient force can be well controlled due to its linear dependence on the input laser power. Here, we propose to exploit the intrinsic nonlinear nature of the optomechanical gradient force to induce a tunable 3-wave coupling between the fundamental modes of two doubly clamped nanophotonic beams. For one of the beams having half the width of the other beam, the 1:2 internal resonance between the fundamental modes supports degenerate spontaneous parametric down-conversion (SPDC). We theoretically explore the main feature of the dissipative phase diagram of the underlying degenerate parametric oscillator model to show that the critical point of the SPDC occurs at parameters which are well in reach of state-of-the-art experiments.
\end{abstract}


\maketitle

\section{Introduction}

In the 1970s Arthur Ashkin and colleagues observed that strong electromagnetic field gradients in close vicinity to dielectric materials induce mechanical forces that result in measurable nanometer-scale displacements. This observation ultimately led to the conceptual development of what is known today as optical tweezers, which have been awarded the Noble prize in 2018~\cite{ashkin1986observation} and inspired optical trapping experiments \cite{bowman2013optical,grzegorczyk2006stable}.
Since then, optically induced gradient forces were widely used to manipulate microparticles, and therefore enabled a multitude of applications in biology; especially in manipulating living cells~\cite{molloy2002lights,kuo1992optical,florin1997photonic,svoboda1994biological,greulich2012micromanipulation}. 

The same effect has been exploited in physics and engineering e.g. for accurate levitation, actuation and assembly of particles and nanostructures~\cite{polimeno2018optical,driscoll2018leveraging,driscoll2018leveraging,romero2011optically,marago2013optical,vcivzmar2010multiple}, and opened the door to the field of optical binding, where microparticles bind to form arrays under the influence of a laser pump \cite{dholakia2010colloquium,burns1989optical,metzger2006visualization}. Optically induced gradient forces also enabled fundamental cavity optomechanical experiments~\cite{aspelmeyer2014cavity} cooling levitated microparticles to sub- and millikelvin temperatures~\cite{kiesel2013cavity,gieseler2012subkelvin,li2013millikelvin}. Even more, electrically induced gradient forces have been employed to actuate and control nanoelectromechanical systems ~\cite{bib:Schmid2008a,bib:Unterreithmeier2009,bib:Rieger2012a,bib:Seitner2017b,leuch2016parametric}.

%
\addPDS{ 
	Optically induced gradient forces in waveguides are scalable to the nanoscale and thus promise a broad range of technological  applications~\cite{li2008harnessing,pernice2009theoretical,quack2019mems,quack2019tue2}. In addition to silicon photonic applications ranging from optical phaseshifters \cite{guo2012broadband,van2012ultracompact,Povinelli:05} to optically-tunable mechanical Kerr coefficients \cite{Ma:11}, and all-optical switches \cite{fan2018novel}, the vision is to exploit the optomechanical gradient forces on a CMOS (complementary metal-oxide-semiconductor)-compatible platform ~\cite{li2008harnessing,li2012multichannel,rochus2017modelling,van2010optomechanical} and enable all optical operation of nano- and micro-electromechanical systems (NEMS/MEMS). Such a novel physical operation principle generates design opportunities to improve in terms of performance, cost and reliability of NEMS and MEMS devices. In particular for sensing applications on the nano-scale optomechanical working principles will have advantages over electrical principles such as for example nano-scale resolution and operation speed~\cite{cai2012nano}.}
	
\addPDS{Micro-electromechanical systems have a longstanding history of academic research~\cite{senturia2007microsystem} and industrial applications~\cite{korvink2010mems}, ranging from sensors such as pressure sensors, gyroscopes and accelerometers to actuators such as micro mirrors~\cite{neul2007micromachined,li2018microactuators,bogue2013recent}. In many cases the capacitive electrostatic forces have been used to actuate and sense the mechanical motion and operate the MEMS device in the linear regime. However, many ideas exploit the fundamental phenomena of nonlinear oscillations to achieve improved device performance~\cite{rugar1991mechanical,fedder2015resonant}, as for example parametrically amplified gyroscopes~\cite{ahn2014encapsulated,nitzan2015self}, or ultrasensitive force sensors~\cite{papariello2016ultrasensitive,karabalin2011signal}.}

\addPDS{Motivated by proof of concept experiments of nano-optomechanical actuation~\cite{van2010optomechanical,cai2012nano}, we show in this paper that, just like NEMS and MEMS in the electrical domain, optomechanically coupled waveguides can also be engineered such that the light forces induce significant nonlinear phenomena for the mechanical modes. In contrast to purely mechanical nonlinearities, the nonlinearities induced by the optomechanical gradient forces are easily tunable via the input laser power during the operation of the device.}

\addPDS{In particular, we discuss a certain geometry which exemplifies the nonlinear mode coupling of two mechanical or rather acoustic modes mediated by light forces. These two mechanical modes with frequency ratio 1:2 describe the fundamental modes of two beams with a 1:2 cross-section ratio. If the wider beam, which acts as a waveguide, is actuated by the optomechanical gradient force, via a modulated optical input power, the induced nonlinear 3-wave coupling between the two mechanical modes leads to the excitation of the non-actuated thinner beam. Such a nonlinear phenomenon is well known as spontaneous parametric down-conversion (SPDC) in the context of nonlinear optics~\cite{boyd2003nonlinear,couteau2018spontaneous}.} 

\addPDS{The methods in our work are generic showing that important concepts widely used in NEMS and MEMS design such as frequency tuning and parametric modulation are within reach of currently available technology. Remarkably, with the typical geometric dimensions, optical gaps and mechanical quality factors the required optical powers do not exceed the milliwatt range.}

\addPDS{The remainder of the paper is organized as follows: We start with a discussion on the basics of optomechanical gradient forces and spontaneous parametric down-conversion in section~\ref{sec:OMGF_SPDC} before section~\ref{sec:Exp} introduces the proposed photonic system and several excitation scenarios in detail. In section~\ref{sec:OR} we propose an optical read-out method to observe the mechanical dynamic behavior by implementing the system into a phase-sensitive optical integrated circuit, a Mach-Zehnder interferometer (MZI). Section~\ref{sec:Methodology} discusses the  theoretical description of the resulting optomechanical system and presents the critical amplitudes for different quality factors and drive detuning parameters of the NS and the WG.}

\section{The optomechanical gradient force and spontaneous parametric down-conversion \label{sec:OMGF_SPDC}}

\begin{figure}[t!]
	\includegraphics[width=\columnwidth]{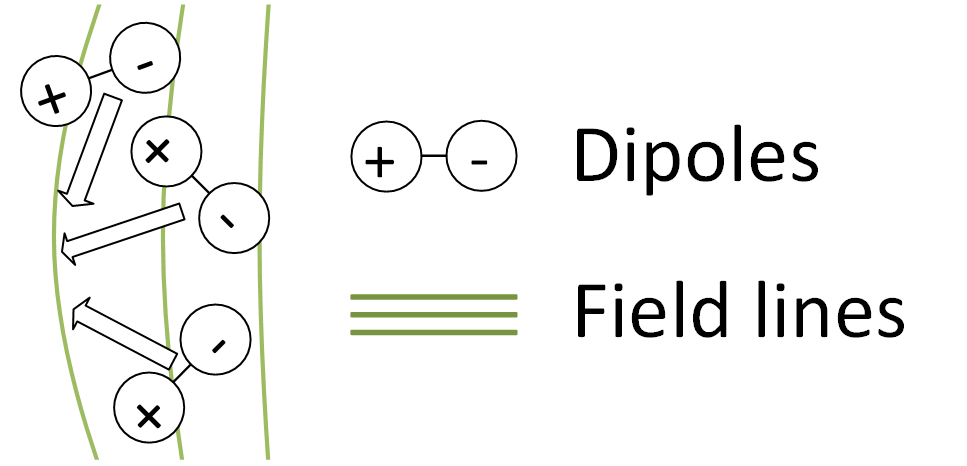}
	\caption{Dipoles accelerated as a result of gradient forces in an inhomogeneous electromagnetic field. \label{fig:Dipoles}}
\end{figure}

Qualitatively, the optically induced gradient force on a dielectric slab exposed to an imhomogeneous electromagnetic field arises from unbalanced forces acting on the oppositely charged ends of induced dipoles. Figure~\ref{fig:Dipoles} illustrates how dipoles are accelerated towards strong field gradient regions. A suspended dielectric nanobeam will thus be mechanically deflected. 
In essence, the resulting optomechanical gradient force mutually couples the mechanical deflection and the optical excitation, and effectively produces an optical element, that responds to changes in optical pump parameters.

The system under our investigation is composed of two suspended dielectric beams of the same height but widths differing by a factor of $2$. The wider beam, referred to as the "waveguide" (WG), supports propagation of an optical mode at the pumping wavelength. The closely spaced thinner beam, referred to as the "nanostring" (NS), does not support any propagating mode at the given wavelength but can be manipulated by the optical pump. For a sufficiently small gap between the WG and the NS, the evanescent portion of the guided mode propagating through the WG will lead to an attractive force pulling the NS towards the WG, given that the excited optical mode propagating through the WG is a symmetric transverse electric mode~\cite{Povinelli:05}.

We theoretically demonstrate a 3-wave coupling between the mechanical modes of the WG and the NS that is promoted and tuned by the optomechanical gradient force. As a result of the 3-wave coupling, the system can no longer be solely described as an optically nonlinear element, but must also include mechanical nonlinearities. In fact, we show that our effective mechanical model resembles the well-known degenerate parametric oscillator (DPO) model in full analogy.


DPOs have been widely studied in the context of nonlinear optics~\cite{boyd2003nonlinear}.  where an optical nonlinear material with $\chi^2$ susceptibility decomposes optically pumped photons at $\omega_0$  into photons having frequencies $\omega_1$ such that $\omega_0 = 2\omega_1$. The conversion, however, only occurs above a certain threshold power of the pumped photon intensity. This phenomenon known as spontaneous parametric down-conversion (SPDC) constitutes a paradigmatic example of a dissipative phase transition~\cite{carmichael2009statistical}. 
SPDC has triggered both technological advances \cite{abadie2011gravitational,takeno2007observation,vahlbruch2008observation}, and fundamental insights in the field of quantum optics \cite{navarrete2008noncritically,walls2008quantum,carmichael2009statistical,degenfeld2015self} ranging from quantum information \cite{braunstein2005quantum,weedbrook2012gaussian,leghtas2015confining,fortsch2013versatile} to modern hybrid optomechanical devices \cite{aspelmeyer2014cavity,benito2016degenerate,degenfeld2016degenerate} and even found use in industrial applications of micro- and nanoelectro-mechanical systems \cite{aubin2004limit,ganesan2017phononic}.

Our system is able to explore the full dissipative steady state phase diagram of the DPO model in a novel optomechanical setup. The WG and the NS are represented by Euler-Bernoulli beams which are actuated by the optomechanical gradient force. The nonlinearity mediating the coupling of the two beams enters via the gap- or rather deflection-dependent force. The controllability of the proposed scheme originates from the proportionality of the optomechanical gradient force on the optical input power. 

\begin{figure*}
	\includegraphics[width=\textwidth]{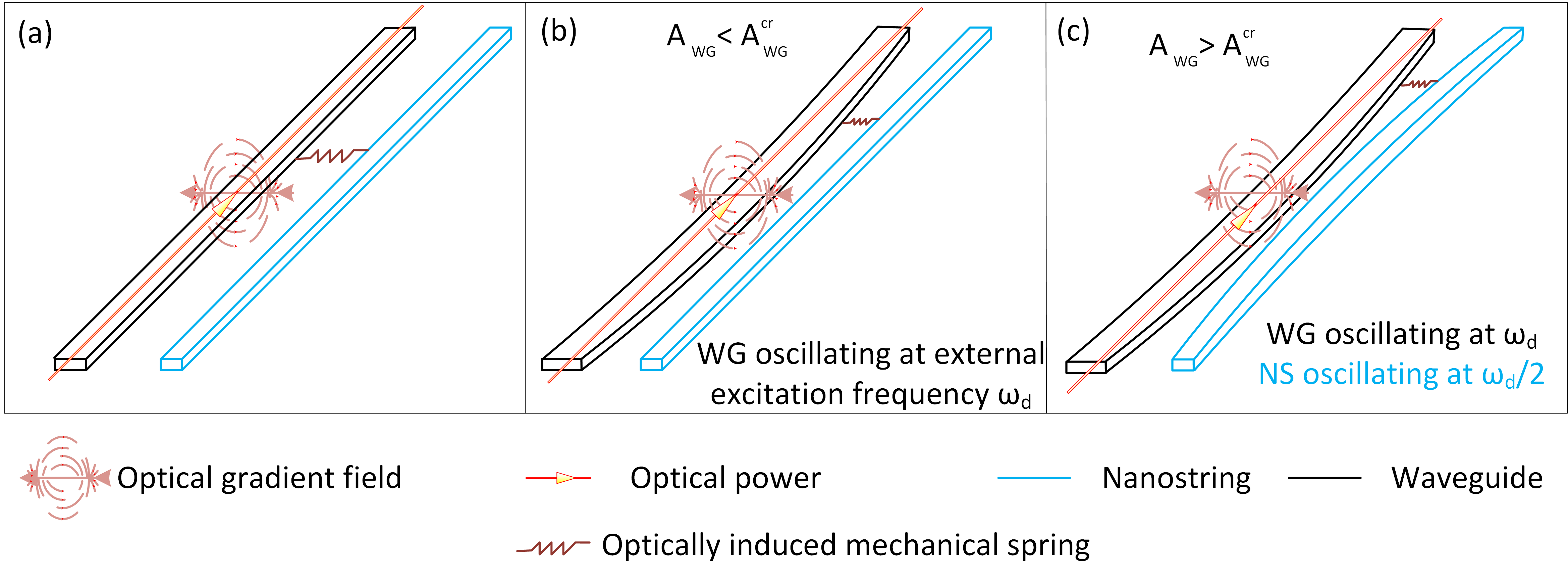}
	\caption{Scheme of the nanophotonic system and its steady-state phases. (a) The waveguide (WG) and the nanostring (NS) connected by a spring symbolizing the optomechanical interaction which mediates the 3-wave coupling required for SPDC. (b) Trivial  phase below the critical oscillation amplitude $A_\text{WG} < A_\text{WG}^\text{cr}$. (c) Non-trivial, symmetry-broken phase above the critical oscillation amplitude $A_\text{WG} > A_\text{WG}^\text{cr}$. Note: Under constant optical pumping both the waveguide and the nanostring exhibit a finite static deflection, which is not included in the figure for the sake of clarity. \label{fig:Experiment}}
\end{figure*}

Figure~\ref{fig:Experiment}(a) schematically depicts the system under investigation. Its steady state phases are illustrated in Figs. ~\ref{fig:Experiment}(b) and (c). In the trivial phase displayed in Fig. ~\ref{fig:Experiment}(b), the fundamental beam mode of the WG oscillates at the drive frequency of an external actuation, while the NS remains mainly at rest. In the non-trivial, symmetry-broken phase shown in Fig. ~\ref{fig:Experiment}(c), the NS oscillates at exactly half of the drive frequency due to the nonlinear 3-wave coupling mediated by the optomechanical gradient force. The critical point of the dissipative phase transition occurs at a critical oscillation amplitude $A_\text{WG}^\text{cr}$ of the WG.  

\section{Waveguide system \label{sec:Exp}}

Previous works have proposed and demonstrated gradient force-based optomechanical interaction in silicon photonics using a waveguide and a parallel nanostring that are released to have an air cladding~\cite{van2010optomechanical,guo2012broadband}. Our work builds upon the same platform. In this section, we introduce the nomenclature as well as the specific geometrical parameters and excitation conditions to be used throughout the paper. 

\begin{figure}[h!]
	\includegraphics[width=\columnwidth]{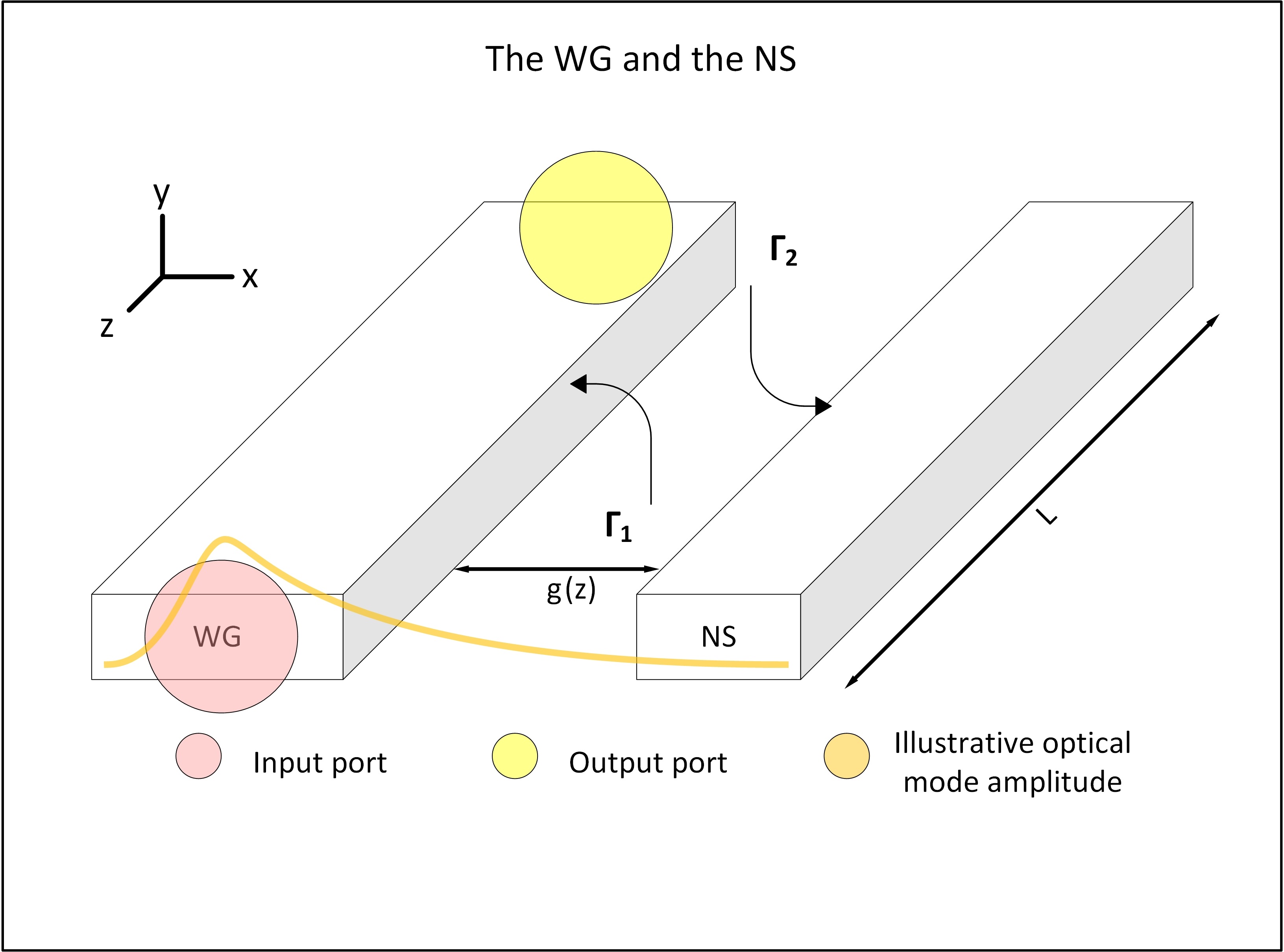}
	\caption{Our system formed of two parallel silicon nano-beams of length L, both of the same height h, the wider is named Waveguide (WG), and the thinner is named Nanostring (NS). $\Gamma_{\text{1,2}}$ symbolize oppositely facing boundaries of the waveguide and the nanostring. g is the variable gap between the boundaries $\Gamma_{\text{1}}$ and  $\Gamma_{\text{2}}$ is defined as a function of the spatial direction z. \label{fig:SlotWaveguide}}
\end{figure}

Our waveguide system is formed of two silicon beams denoted WG and NS of the same height h and length L as illustrated in Fig.~\ref{fig:SlotWaveguide}. Waveguide (WG) refers to the wider beam of width $W_{\text{WG}}$, and nanostring (NS) denotes the thinner beam  of width $W_{\text{NS}}$. The gap $g$ refers to the distance between two opposite points along the waveguide and the nanostring. Initially at zero optical power, the gap has a value of $G_{\text{0}}$. Table \ref{tab:GEOParam} lists the geometric parameters of the WG and the NS. 

\begin{table}[h]
	\caption{\label{tab:GEOParam}Geometric parameters of the WG and the NS.}
	\begin{ruledtabular}
		\begin{tabular}{lllll}
			L ($\mu$m) & $G_{0}$ (nm) & $W_{\text{WG}}$ (nm) & $W_{\text{NS}}$ (nm) & $h$ (nm) \\
			100 & 130 & 450 & 225 & 225\\
		\end{tabular}
	\end{ruledtabular}
\end{table}

The optical excitation is done through the input port of the waveguide as also illustrated in Fig.~\ref{fig:SlotWaveguide}. In practice, the waveguide inputs and outputs can be connected to optical grating couplers for optical excitation and read-out. 

We consider three excitation and pumping conditions:
1) The WG is only pumped by an unmodulated optical power. 2) The WG is optically pumped with a constant unmodulated optical power, and simultaneously excited by a mechanical shaker force with a drive frequency $\omega_{d}\approx\omega_{\text{WG}}$ 3) The WG is optically excited with an amplitude modulated optical power, that is superimposed over a constant biasing optical power, with a modulation frequency $\omega_{\text{d}}$ $\approx$ $\omega_{\text{WG}}$, and no direct mechanical driving.

\section{Optical read-out of mechanical response \label{sec:OR}}

The optical input power injected into the WG determines the gradient force and thus the gap to the NS. When the waveguide system consisting of the WG and the NS is inserted in one side of an optical Mach-Zehnder-Interferometer (MZI), a change of the size of the gap will affect the the excited optical mode refractive index. The output of the MZI, therefore, presents the optically tuned phase and transmission response of the waveguide. This was experimentally verified for a silicon nitride-based system having a WG and a NS suspended parallel to each other~\cite{Fong:11}. The work targeted and demonstrated optical phase measurements, thus showing the possibility of a tunable optical device based on optomechanical gradient forces in such a system.

\begin{figure}
	\includegraphics[width=\columnwidth]{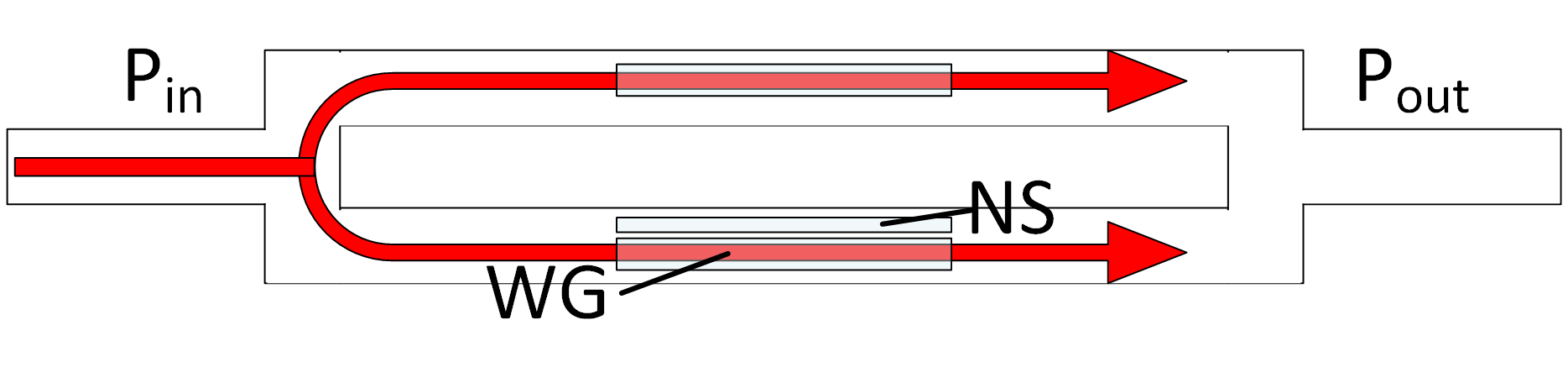}
	\caption{Mach-Zehnder Interferometer with one of its sides having the proposed optomechanical system of the WG and the NS.\label{fig:OMTMZI}}
\end{figure}

We propose to exploit the same idea to read out the mechanical response of the waveguide system: Both the WG and the NS are suspended parallel to each other in one arm of the MZI as shown in Fig.~\ref{fig:OMTMZI}. The optical phase-dependence of the output optical signal as a function of the input optical power is recorded. By formulating the optical transmission in terms of the optical phaseshift, we directly correlate the mechanical deflection to the optical phaseshift and transmission signals to provide a read-out method for the proposed excitation schemes.

The optical phase response can be defined in terms of the input optical power in the proposed system (See appendix \ref{App:OpticalReadout} for detailed derivation) by

 \begin{subequations}
\label{eqn:Theory:MZI_5}
\begin{align}
\delta \phi = \frac{2 \pi}{\lambda} (\int_{\text{0}}^{\text{L}} n_{\text{eff}}(G_{\text{0}}-(u_{\text{WG}}(z)+u_{\text{NS}}(z)))dz - \nonumber \\ n_{\text{eff}}(G_{\text{0}})L),
\tag{\ref{eqn:Theory:MZI_5}}
\end{align}
\end{subequations} 

where $n_{\text{eff}}$ is the function relating the effective optical refractive index of the WG and the NS to the gap in-between. This will be further discussed in section~\ref{sec:Methodology}.

The resulting optical transmission as a function of the optical phaseshift is defined by 

\begin{equation}
T = \frac{P_{\text{out}}}{P_{\text{in}}} = \frac{1}{2}(1+\cos(\delta \phi(P_{\text{in}}) - \delta_{\text{imbalance}})).
\label{eqn:Theoryi:MZI_10}
\end{equation}


Therefore, the optical phaseshift is a direct function of the input power, and consequently the optical transmission. We later predict the optical phaseshift response for the proposed excitation cases, which can be inferred from experimental measurements of optical transmission using eq.~(\ref{eqn:Theoryi:MZI_10}). Note that there is an initial optical imbalance between the two optical paths of the MZI due to the possibly different initial optical refractive indices; we denote this imbalance by $\delta_{\text{imbalance}}$, which we set to zero without loss of generality. 


\section{Optomechanical model \label{sec:Methodology}}

The optomechanical gradient force is defined as~\cite{Povinelli:05}

\begin{equation}
F_\text{OMG}(t) = - \frac{LP(t)}{c} \frac{\partial n_{\text{eff}}}{\partial g} ,
\label{eqn:FOMG}
\end{equation}

where $P$ is the optical power, which can be in general time-dependent, $n_{\text{eff}}$ is the total effective optical refractive index of a guided propagating optical mode promoting the gradient field through its evanescent tail, L is the length of the dielectric nanostring exposed to the evanescent field, c is the speed of light in vacuum, and g is the z-dependent distance from boundary $\Gamma_{\text{1}}$ of the waveguide to boundary $\Gamma_{\text{2}}$ of the nanostring, see also Fig.~\ref{fig:SlotWaveguide}.

The deflection resulting from a gradient optomechanical force in a system of a NS suspended parallel to a suspended WG is governed by coupling the Euler-Bernoulli beam equation with the optomechanical gradient force introduced in eq.~(\ref{eqn:FOMG}). This yields

\begin{subequations}
	\label{eq:EBTD}	
	\begin{align}
	\frac{F_{\text{OMG}}}{L}(G_{\text{0}}-({u_{\text{WG}}}(z,t)+{u_{\text{NS}}}(z,t))) = \nonumber \\ \rho A_{\text{i}} \ddot{u}_{\text{i}}(z,t) + d_{\text{i}} \dot{u}_{\text{i}}(z,t) + \frac{YA_{\text{i}}{W_{\text{i}}}^{\text{2}}}{12 L^{\text{4}}}{{u_{\text{i}}}^{''''}}(z,t) - \nonumber \\ (\frac{T_{\text{0}}}{L^{\text{2}}} + \frac{\Delta T_{\text{i}}}{L^{\text{2}}}){{u_{\text{i}}}^{''}}(z,t) \tag{\ref{eq:EBTD}},
	\end{align}
\end{subequations}

where $\rho$ represents the material's density, $A_{\text{i}}$ the cross-sectional area of the beam, $d_{\text{i}}$ the mechanical damping per length, Y the Young's modulus, h and L the height and the length of the beam respectively, and $u_{\text{i}}(z,t)$ is the time-dependent absolute spatial deflection of the beams along the z-axis, where i=\{WG,NS\}.

We have introduced eq.~(\ref{eq:EBTD}) such that the coordinate $z \in [0,1]$ denotes the dimensionless length and the dot and prime denote the derivative with respect to time and the dimensionless length $z$, respectively. Moreover, we consider the case of clamped-clamped boundary conditions with $u_{\text{i}}(0,t) = u^{'}_{\text{i}}(0,t) = u_{\text{i}}(L,t) = u^{'}_{\text{i}}(L,t) = 0$. In eq.~(\ref{eq:EBTD}) we keep the sign of the force positive for both the WG and the NS. The attractive nature of this force is accounted for by using the time dependent gap definition $g(z,t) = G_{\text{0}}-({u_{\text{WG}}}(z,t)+{u_{\text{NS}}}(z,t))$, which reflects a decreasing gap value as the optical power increases.

It is essential to note that the term $\Delta T$ representing the bending tension is the main source of geometric nonlinearity (GNL) with

\begin{equation}
\begin{gathered}
\Delta T_{\text{i}} = \frac{YA_{\text{i}}}{2L^{\text{2}}} \int_{\text{0}}^{\text{1}} ({{u^{'}_{\text{i}}}}(z,t))^2 dx.
\label{eq:bm}
\end{gathered}
\end{equation}

We also consider an initial stress-free material with $T_{\text{0}}=0$ in our analysis. Relaxing this assumption, however, will not have an impact on the following proposal.

The full transient deflection described by eq.~(\ref{eq:EBTD}) can be numerically solved. However, this approach is computationally expensive and the essential physics remains mostly illusive. In this work, we rewrite the equations of motion in the basis of the linear normal modes of the beams~\cite{bokaian1990natural} and approximate the overall mechanical deflections by the fundamental modes of the WG and NS only. Within our approximation we are, however, still able to account for the relevant nonlinearities. 

In the following subsections, we will perform 1) A static solution, that sweeps across time-invariant optical power values to obtain the deflection of the WG and the NS from the time-invariant Euler-Bernoulli equation. This static solution allows us to define the regions of stable device operation and determine the bias points which ultimately define the coupling strength of the beams induced by the OMG forces. We show that our modal reduction scheme can accurately determine the instability point of the nonlinear system by verifying the results with nonlinear finite element static simulations. 2) A dynamic analysis around the static deflection point of the beams. Here, we illustrate how our system resembles the well-known model of degenerate optical parametric oscillators~\cite{davies2008statistical,degenfeld2015self} with a tunable 3-wave coupling strength between the fundamental modes of the waveguide and nanostring. 

\subsection{Gap-dependent refractive index}
The first step of our analysis requires to determine the optomechanical gradient force as a function of the gap. Therefore, we obtain the total effective refractive index comprising the effective optical mode of the guided part in the WG and the evanescent component interacting with the NS from an optical mode simulator. The obtained refractive index is then fitted against the gap between the NS and the WG. We found that the gap dependent refractive index function can be well fitted to

\begin{subequations}
	\label{eqn:LumericalNeff}
	\begin{align}
	n_{\text{eff}}(G_{\text{0}} - U_{\text{t}}(z)) = a_{\text{o}}e^{(b_{\text{o}} (G_{\text{o}} - U_{\text{t}}(z))) } + c_{\text{o}}e^{(d_{\text{o}} (G_{\text{o}} - U_{\text{t}}(z)))},
	\tag{\ref{eqn:LumericalNeff}}
	\end{align}
\end{subequations}

where the fitting parameters $a_{\text{0}}, b_{\text{0}}, c_{\text{0}}$ and $d_{\text{0}}$ are dependent on the width and the height of the NS and the WG, but they are independent of the waveguide length or the input optical power, and $U_{\text{t}}(z) = u_{\text{WG}}(z)+u_{\text{NS}}(z)$ as the total deflection magnitude of both the WG and the NS. For obtaining these parameters the c-band center wavelength of 1550 nm was used, and we employed Lumerical Mode - a commercial software for obtaining refractive indices and profiles of optical modes - to obtain the refractive index for the specific geometry summarized in table~\ref{tab:GEOParam}.

Within our fit function the parameters $a_{\text{0}}$, and $c_{\text{0}}$ turn out to be positive constants, while the parameters $b_{\text{0}}$, and $d_{\text{0}}$ are negative constants. This reflects a negative slope of the fit function as the gap value decreases, or as the deflection of the waveguide and the nanostring increases due to their mutual attraction.

\subsection{Modal projection\label{subsec:MODALProjection}}

The field variable $u_{\text{i}}(z,t)$ in the Euler-Bernoulli equation denotes the deflection at every material point of the WG and NS and therefore varies along the z-direction due to the clamped boundaries and the WG and the NS being subjected to a lateral optomechanical force. In the following, we will determine the dimensionless modal shape functions $S_{\text{i,n}}(z)$ with i=\{WG,NS\} which form an orthonormal basis and therefore allow us to introduce the modal representation of the deflection, given by

\begin{equation}
	u_{\text{i}}(z,t)=\sum_{\text{n}}^{\infty} q_{\text{i,n}}(t) S_{\text{i,n}}(z). \label{eq:Modeshape}
\end{equation}

The modal amplitudes are formally defined by 

\begin{equation}
	q_{i,n} \overset{\mathrm{def}}{=} \int_{\text{0}}^{\text{1}}dz S_{\text{i,n}}(z)u_{\text{i}}(z,t).
\end{equation}

The shape functions, for which we use the convention that $S_{\text{WG,n}}(z)=-S_{\text{NS,n}}(z)$, since they are identical for the WG and the NS but of opposite sign, are determined by the eigenvalue problem   

\begin{equation}
\frac{YA_{\text{i}}{W_{\text{i}}}^{\text{2}}}{12L^{\text{4}}} {S^{''''}_{\text{i,n}}}(z) = \omega_{\text{i,n}}^2 \rho A_{\text{i}} S_{\text{i,n}}(z),
\label{eq:EBM}
\end{equation}

with the boundary conditions $S_{\text{i,n}}(1) = S_{\text{i,n}}(0) = S^{'}_{\text{i,n}}(1) = S_{\text{i,n}}^{'}(0)=0$ for the clamped-clamped case for all values of $n$.  The integer $n$ denotes the mode number and the eigenvalues $\omega_{\text{i,n}}=r_{\text{n}}^{\text{2}}\frac{W_{\text{i}}}{2\pi L^2}\sqrt{\frac{Y}{12\rho}}$ for $n \in \{1,2,3,...\}$ the corresponding angular resonance frequencies of the modes, with the roots $r_{\text{n}}$ determined from the transcendental equation $\cosh(r_{\text{n}})\cos(r_{\text{n}})-1=0$. The roots and consequently the angular frequencies are monotonically increasing with $r_{\text{1}}\approx4.7$, $r_{\text{2}}\approx7.9$, $r_{\text{3}}\approx11$, ... .

The closed form of the modal shape functions are specified in Appendix~\ref{App:EBP}. Note that the angular frequencies scale linearly with the width of the beams in the bending direction, and therefore in the ideal design case we have $\omega_{\text{WG,1}}=2\omega_{\text{NS,1}}$.

In order to rewrite the Euler Bernoulli equation, eq.~(\ref{eq:EBTD}), into its modal representation for the n-th modal amplitude $q_{\text{i,n}}$(t), employ a prodecure referred to as modal projection: We insert eq.~(\ref{eq:Modeshape}) into eq.~(\ref{eq:EBTD}), multiply by the linear normal mode shape $S_{\text{n}}(z)$ and then integrate the equation over the length of the beams according to $\int_{\text{0}}^{\text{1}} Ldz$. We find 

\begin{subequations}
	\label{eq:modaleqs}
	\begin{align}
		{F_{\text{OMG,n}}^{\text{i}}}({q}_{\text{WG,n}},{q}_{\text{NS,n}}) = m_{\text{i,n}}\ddot{q}_{\text{i,n}}(t) + d_{\text{i,n}}\dot{q}_{\text{i,n}}(t) + \nonumber \\ K_{\text{i,n}}q_{\text{i,n}}(t) + \sum_{\text{m,l,k=1}}^{\infty} \beta_{\text{n,m,l,k}}^{\text{i,GNL}}q_{\text{i,m}}(t)q_{\text{i,l}}(t)q_{\text{i,k}}(t),
		\tag{\ref{eq:modaleqs}}
	\end{align}	
\end{subequations}

with the modal mass, modal damping, modal stiffness, modal amplitude dependent modal optomechanical gradient force, and modal 4-wave couplings induced by the geometric nonlinearity denoted by $m_{\text{i,n}}$, $d_{\text{i,n}}$, $K_{\text{i,n}}$, $F_{\text{OMG,n}}^{\text{i}}$, and $\beta_{\text{n,m,l,k}}^{\text{i,GNL}}$, respectively. The exact definitions of these quantities can be found in Appendix~\ref{App:EBP}. The modal stiffness relates to the angular frequency and modal mass as $K_{\text{i,n}}=m_{\text{i,n}} \omega_{\text{i,n}}^{\text{2}}$. 

The modal representation of the Euler-Bernoulli equation is exact. However, it requires to solve an ordinary differential equation for infinitely many modal amplitudes if no further approximations are made. In contrast to the Euler-Bernoulli equation in its continuous form, eq.~(\ref{eq:EBTD}), the modal representation allows us to perform a drastic complexity reduction by approximating the dynamics only with the fundamental modes of the beams. Thus, we set $q_{\text{i,n}}(t)=0$ for all $n>1$ and therefore simplify the deflection by $u_{\text{i}}(z,t)\approx q_{\text{i,1}}(t) s_{\text{1}}(z)$. This ultimately reduces the complexity to an ordinary differential equation with only two variables. Moreover, the optomechanical gradient force reduces to $F_{\text{OMG,1}}^{\text{WG}}(q_{\text{WG,1}}+q_{\text{NS,1}},t)=-F_{\text{OMG,1}}^{\text{NS}}(q_{\text{WG,1}}+q_{\text{NS,1}},t)$, where $F^\text{i}_\text{OMG,1}(q_{\text{WG,1}}+q_{\text{NS,1}})$ is the time-dependent modal projection of the optomechanical gradient force in the modal domain using the fundamental mode of the Euler-Bernoulli clamped-clamped beam. The derivation of the static form of $F_\text{OMG,1}^\text{i}$ and the need of a modal force representation is detailed in the following section.

\subsection{Static analysis}

As we will elaborate on in the following, one of the key ingredients of our proposal is the tunable 3-wave coupling strength between the two beams. In fact, the 3-wave coupling is linearly proportional to the static input laser power $P_\text{dc}$. However, the input power is upper bounded by a certain threshold at which our system will collapse due to the attractive nature of the optomechanical gradient force. The goal of this section is to determine this threshold. i.e. the snapping power and snapping deflection to identify the regime of safe device operation. 

One possibility to find the snapping power is to substitute eq.~(\ref{eqn:LumericalNeff}) into eq.~(\ref{eqn:FOMG}) and solve for the static spatial deflection with $\dot{u}(z,t)=\ddot{u}(z,t)=0$ according to the static Euler-Bernoulli equation

\begin{subequations}
	\label{eqn:EulerBernoulliAndNeff}	
	\begin{align}
		\frac{YA_{i}h^2}{12 L^4} \frac{\partial^4 u_{i}(z)}{\partial z^4} = \frac{P_{dc}}{c} *  
		(a_{o}b_{o}e^{(b_{o} (G_{o} - U_{t}(z))) } + \nonumber \\
		 c_{o}d_{o}e^{(d_{o} (G_{o} - U_{t}(z)))}).
		\tag{\ref{eqn:EulerBernoulliAndNeff}}
	\end{align}
\end{subequations}

The exponentials on the right hand side of eq.~(\ref{eqn:EulerBernoulliAndNeff}) render the equation as a partial differential equation of infinite order. This hinders obtaining an analytic solution without further approximations. One possible approximation involves the linearization of the exponential around a gap value of interest \cite{Povinelli:05,guo2012broadband}. For large beam deflections or rather large input powers, however, fully numerical Finite Element (FE) approaches have been proposed \cite{ozer2018stability}. We have also implemented a FE model, which we detail in Appendix~\ref{App:FEM}, to verify our modal approximation \cite{ashour2019stability}.

Within the modal approximation, we can fit the optomechanical gradient force to 

\begin{subequations}
	\label{eqn:FOMGmtot}
	\begin{align}
		F_{OMG,1}^{i}(q_{tot})=-\frac{LP_{dc}}{c}(c_{1}e^{c_{\text{2}}(G_{0}-q_{tot})}+ \nonumber \\ c_{3}e^{c_{\text{4}}(G_{0}-q_{tot})}) \tag{\ref{eqn:FOMGmtot}}
	\end{align}
\end{subequations}

with width and height dependent negative fit coefficients $c_{1}, c_{2}, c_{3}$, and $c_{4}$, and the total modal amplitude $q_{tot}=q_{NS,1}+q_{WG,1}$. The modal equations, see eq.~(\ref{eqn:EulerBernoulliAndNeff}), simplify to 

\begin{subequations}
	\label{eqn:KEq}
	\begin{align}
		F^{i}_{OMG,1}(q_{tot}) = K_{i,1}q_{i,1} + \beta^{i,GNL}_{1,1,1,1}q_{i,1}^{3}  \tag{\ref{eqn:KEq}}
	\end{align}
\end{subequations}

\subsubsection{Maximum nonlinear static deflection\label{subsec:SSD}}

We define the physically meaningful static stability by the maximum deflection values corresponding to $P_{\text{dc}}$ where the mechanical restoration forces can balance the attractive optically induced mechanical force. This can be represented by defining an effective stiffness matrix of the whole mechanical system including the WG, the NS, and the optical gradient force using 

\begin{equation}
K_{\text{stiffness}} = K_{\text{GNL}} - K_{\text{OMG}},
\label{eqn:statstab}
\end{equation}

with equations 

\begin{equation}
(K_\text{OMG})_\text{i,j} = \frac{\partial F_\text{OMG,1}^\text{i}}{\partial q_{j,1}}\bigg|_{q_\text{tot}=q_\text{sol}(P_\text{in})},
\label{eqn:komg}
\end{equation}

and 

\begin{equation}
	(K_\text{GNL})_\text{i,j} = \frac{K_\text{i}q_\text{i,1}+\beta_\text{NS}q^\text{3}_\text{i,1}}{\partial q_\text{j,1}}\bigg|_{q_\text{tot}=q_\text{sol}(P_\text{in})},
	\label{eqn:kngl}
\end{equation}

representing the stiffness matrices of the optomechanical gradient force and the geometric nonlinearity of the WG and the NS, respectively. Moreover, $q_{\text{sol}}$ is given by the solution of eqs.~(\ref{eqn:FOMGmtot}) and (\ref{eqn:KEq}) for different optical powers.

The eigenvalues of $K_{\text{stiffness}}$ have negative values. They are non-physical and represent unstable solutions. In other words, negative eigenvalues indicate an optically induced attraction force exceeding the mechanical stiffness forces of the WG and the NS and which leads to the snapping of the waveguides. Consequently, the snapping power as well as the snapping deflections are determined by the first appearance of a negative eigenvalue of the stiffness matrix. 

\subsubsection{Static analysis results\label{subsec:SAR}}

\begin{figure*}
	\centering
	\includegraphics[width=1\textwidth]{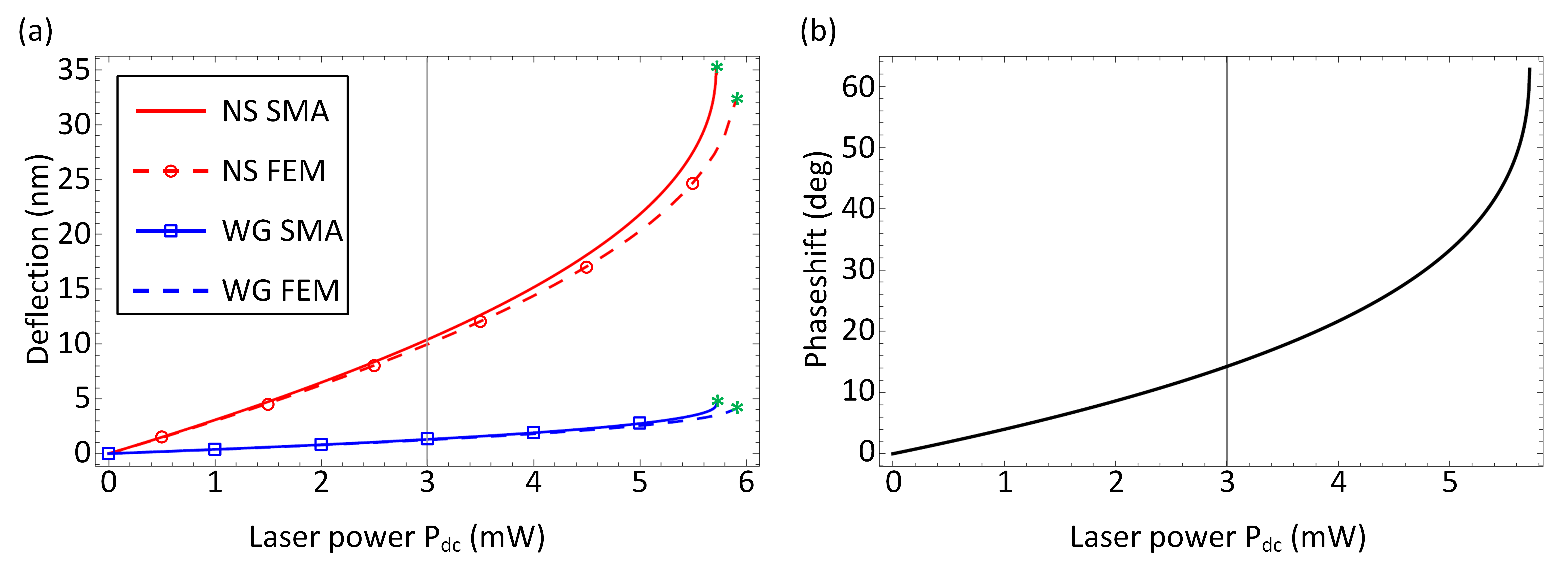}
	\caption{(a) Nanostring and waveguide deflections vs. constant input laser power from the single mode approach (SMA) shown by the solid red and blue lines, respectively, and the FE analysis shown by the dashed red and blue lines, respectively. (b) Optical phaseshift vs. constant input laser power. The snapping power is found to be at ${{P^{snap}_\text{dc}}} \approx 5.8~\text{mW}$. The green asterisk marks the non-analytical points in both the SMA as well as the FEM solutions. The gray vertical line at $P_\text{dc}$ = 3 mW marks the bias laser power around which we perform the dynamic analysis.\label{fig:NSNSFF}}
\end{figure*}

Figure~\ref{fig:NSNSFF}a plots the deflection of the WG and the NS as a function of the input optical power $P_\text{dc}$, while Fig.~\ref{fig:NSNSFF}(b) displays the reachable optical phaseshift. 

The solid and dashed curves show a very good agreement between the modal approach and the FE method, validating the description of the essential nonlinear behavior with the fundamental modes only. The deflection calculated using FE has been evaluated at the node lying exactly in the middle of the beams to be comparable to the modal amplitudes $q_\text{WG,1}$, and $q_\text{NS,1}$. However, we have also verified that the overall beam deflection shapes from the static FEM result coincide almost perfectly with the fundamental mode shapes. 

The non-analytic points in Fig.~\ref{fig:NSNSFF}(a) mark the snapping power  ${P^{snap}_\text{dc}}\approx5.8~\text{mW}$ and correspondingly the snapping amplitudes ${q^{snap}_\text{WG,1}}\approx5~\text{nm}$ and ${q^{snap}_\text{NS,1}}\approx35~\text{nm}$. Below these points the system will remain stable. Moreover, the modal approximation will accurately describe the system which we exploit further for the dynamic analysis in the next section. 

\subsection{Dynamic analysis}

Within our dynamic analysis we first state the equations of motions in their modal representation for both the waveguide and the nanostring in subsection~\ref{subsec:EOM}. In subsection~\ref{subsec:ROW} we introduce a Taylor expansion of the exponential function in eq.~(\ref{eqn:FOMGmtot}) to account for nonlinearities up to the third order of the optomechanical gradient force, and thus for the whole optomechanical system. We then apply a rotating wave approximation to the equations of motions, to illustrate the equivalence of our model to degenerate parametric oscillators, and calculate the critical points and loads for the dissipative phase transition for the relevant parameter ranges of our system.  

\subsubsection{Equations of motion \label{subsec:EOM}}

The equation of motions for the WG and NS are introduced in their modal representation to read

\begin{subequations}
	\label{eqn:Fdyn_mechWG}	
	\begin{align}
	F^{\text{WG}}_{\text{OMG,1}}(q_{\text{tot}}) + F_{\text{vib}}=	m_{\text{WG,1}} \ddot{q}_{\text{WG,1}} + \nonumber \\ \frac{m_{\text{WG,1}} \omega_{\text{WG,1}}}{Q_{\text{WG}}} \dot{q}_{\text{WG,1}} + m_{\text{WG,1}} \omega_{\text{WG,1}}^2 q_{\text{WG,1}} +  \nonumber \\ \beta_{\text{1,1,1,1}}^{\text{WG,GNL}} q_{\text{WG,1}}^3,
		\tag{\ref{eqn:Fdyn_mechWG}}
	\end{align}

\end{subequations}

\begin{subequations}
	\label{eqn:Fdyn_mechNS}	
	\begin{align}
		F^\text{NS}_\text{OMG,1}(q_\text{tot}) = m_\text{NS,1} \ddot{q}_\text{NS,1} + \frac{m_\text{NS,1} \omega_\text{NS,1}}{Q_\text{NS}}\dot{q}_\text{NS,1}  + \nonumber \\  m_\text{NS,1} \omega_\text{NS,1}^2 q_\text{NS,1} +   \beta_\text{1,1,1,1}^\text{NS,GNL} q_\text{NS,1}^3, \tag{\ref{eqn:Fdyn_mechNS}}
	\end{align}

\end{subequations}

where $Q_\text{WG}$ and  $Q_\text{NS}$ denote the quality factors of the fundamental modes of the WG and the NS, respectively. The external vibratory force at the drive frequency $\omega_\text{d}$ is denoted by $F_{\text{vib}}$, which we only include in the equation of the WG since it is highly off-resonant for the NS. 

We then use a Taylor expansion to expand the optomechanical force, eq.~\ref{eqn:FOMGmtot}, up to third order keeping only resonant terms, thus accounting for 3-wave, Kerr and cross-Kerr nonlinearities in

\begin{subequations}
	\label{eqn:ModWGDyn}	
	\begin{align}
		F^\text{WG}_\text{OMG,1}(q_{\text{WG,1}},q_{\text{NS,1}}) \approx LP_{\text{dc}}\big(g_\text{0} +  g_\text{1} q_\text{WG,1}(t) + \nonumber \\ g_\text{2} q_\text{NS,1}^2(t) +  
		g_\text{3}q_\text{WG,1}^\text{3}(t) + 3g_\text{3}q_\text{NS,1}^2(t) q_\text{WG,1}(t)\big), \tag{\ref{eqn:ModWGDyn}}
	\end{align}

\end{subequations}

and

\begin{subequations}
	\label{eqn:ModNSDyn}	
	\begin{align}
		F^\text{NS}_\text{OMG,1}(q_\text{WG,1},q_\text{NS,1}) \approx LP_\text{dc}\big(g_\text{0} + g_\text{1} q_\text{NS,1}(t) + \nonumber \\ 2g_\text{2} q_\text{NS,1}(t) q_\text{WG,1}(t) + g_\text{3}q_\text{NS,1}^3(t) + \nonumber \\ 3g_\text{3}q_\text{WG,1}^2(t) q_\text{NS,1}(t)\big),
		\tag{\ref{eqn:ModNSDyn}}
	\end{align}

\end{subequations}

where the parameters $g_\text{0}$, $g_\text{1}$, $g_\text{2}$, and $g_\text{3}$ represent the Taylor coefficients of the projection of $F_\text{OMG}$ upon the fundamental mode shapes of the clamped-clamped beams. The quantity $P_\text{dc}Lg_\text{2}$ is of utter importance in this analysis, as it is the 3-wave coupling strength or rather the analog of the down-conversion rate which induces degenerate spontaneous parametric down-conversion between the fundamental modes of the two beams. Once again, we stress that this coupling is mediated by the optomechanical gradient force and as such depends on the geometry parameters of the system and most importantly it scales linearly with the optical input power $P_\text{dc}$. Therefore, in contrast to the down-conversion rate in nonlinear optics which is proportional to the crystal's nonlinear susceptibility, it is tunable via the bias optical power.

\subsubsection{Rotating wave approximation\label{subsec:ROW}}

We use a rotating wave approximation to calculate the stationary or rather steady state amplitudes of the WG and the NS. We start with defining the modal amplitudes of oscillations in terms of complex normal coordinates $a_\text{WG}$ and $a_\text{NS}$ and the drive angular frequency $\omega_\text{d}$, by \cite{nabholz2019spontaneous}

\begin{subequations}
	\label{eqn:QWG}	
	\begin{align}
		q_\text{WG,1} = \sqrt{\frac{\hbar}{2\omega_{\text{WG,1}} m_{\text{WG,1}}}} \big(a_{\text{WG}}(t)e^{-i\omega_{\text{d}}t}
		+ a_{\text{WG}}^{*}(t) e^{i\omega_{\text{d}} t}\big), \tag{\ref{eqn:QWG}}
	\end{align}

\end{subequations}

\begin{subequations}
	\label{eqn:QNS}	
	\begin{align}
		q_\text{NS,1} = \sqrt{\frac{\hbar}{2\omega_{\text{NS,1}} m_{\text{NS,1}}}} \big(a_{\text{NS}}(t)e^{-i\frac{\omega_{\text{d}}}{\text{2}}t} + a_{\text{NS}}^{\text{*}}(t) e^{i\frac{\omega_{d}}{2} t}\big). \tag{\ref{eqn:QNS}}
	\end{align}
\end{subequations}
 
Following standard steps~\cite{nabholz2019spontaneous}, we derive the equations of motion for the normal coordinates given by
 
\begin{equation}
	 \dot{a}_{\text{WG}} =
	 (-i \delta_{\text{WG}} - d^\text{RWA}_{\text{WG}})a_{\text{WG}} - i \epsilon^{\text{*}} - i \frac{\alpha}{2} a_{\text{NS}}^{\text{2}},
	 \label{eqn:AWGreal}
\end{equation}
\begin{equation}
	 \dot{a}_{\text{WG}}^{*} =
	 (i \delta_{\text{WG}} - d^\text{RWA}_{\text{WG}})a^{\text{*}}_{\text{WG}} + i \epsilon+ i \frac{\alpha}{2} (a^{\text{*}}_{\text{NS}})^{\text{2}},
	 \label{eqn:AWGcomplex}
\end{equation}
\begin{equation}
	 \dot{a}_{\text{NS}} =
	 (-i \delta_{\text{NS}} - d^\text{RWA}_{\text{NS}})a_{\text{NS}} - i \alpha a_{\text{WG}} a^{\text{*}}_{\text{NS}},
	 \label{eqn:ANSreal}
\end{equation}
\begin{equation}
	\dot{a}_{\text{NS}}^{\text{*}} =
	(i \delta_{\text{NS}} - d^\text{RWA}_{\text{NS}})a_{\text{NS}}^{\text{*}} + i \alpha a_{\text{WG}}^{\text{*}} a_{\text{NS}},
	\label{eqn:ANScomplex}
\end{equation}

where the 3-wave coupling term $\alpha$ of the NS and the WG promoted by $F_\text{OMG}$ is found to be linearly proportional to the input optical power amplitude as

\begin{equation}
	\alpha := \frac{-P_{\text{dc}}L g_{\text{2}}\sqrt{\hbar}}{\sqrt{2 \omega_{\text{WG,1}} \omega_{\text{NS,1}}^2 m_{\text{WG,1}} m_{\text{NS,1}}^{\text{2}}}}.
	\label{eqn:ROW3Wave}
\end{equation}

The external excitation term $\epsilon$ is defined in the case of an external vibratory excitation at angular frequency $\omega_d$ as  

\begin{equation}
\epsilon = \frac{i0.52 \rho L A_\text{WG} a_\text{vib}}{\sqrt{8\hbar \omega_\text{WG,1} m_\text{WG,1}}},
\label{eqn:ROWForcePiezo}
\end{equation}

and for an externally modulated optical power at the angular frequency $\omega_d$ as

\begin{equation}
\epsilon = \frac{iP_\text{ac} L g_\text{0}}{\sqrt{8\hbar \omega_\text{WG,1} m_\text{WG,1}}},
\label{eqn:ROWForceOptical}
\end{equation} 

where $P_\text{ac}$ denotes modulated optical power.

As we excite the WG either with a mechanical shaker or a modulated optical input power of angular frequency $\omega_\text{d}$, the mechanical oscillations will nonlinearly detune the frequencies of the fundamenteal modes of the WG and the NS from the drive frequency and its half-wave component, respectively. We model this detuning for the WG and the NS by 

\begin{subequations}
	\label{eqn:detuningWG}	
	\begin{align}
		\delta_\text{WG} := \omega_\text{WG,1} - \omega_\text{d} + \delta \omega^\text{OMG}_\text{WG}  +  \delta \beta^\text{WG} + \delta V_\text{NS},
		\tag{\ref{eqn:detuningWG}}
	\end{align}

\end{subequations}

\begin{subequations}
	\label{eqn:detuningNS}
	\begin{align}
		\delta_\text{NS} := \omega_\text{NS,1} - \omega_\text{d} / 2 + \delta \omega^\text{OMG}_\text{NS}  + \delta \beta^\text{NS} + \delta V_\text{WG}.
		\tag{\ref{eqn:detuningNS}}
	\end{align}

\end{subequations}

We account for three factors that contribute to the frequency detuning. First the linear frequency shift due to the optomechanical gradient force power, that is independent of the oscillation amplitude. We denote this term by

\begin{equation}
\begin{gathered}
	\delta \omega_\text{i}^\text{OMG} := -\frac{P_\text{dc}L g_\text{1} }{2 \omega_\text{i,1} m_\text{i,1}}.
\end{gathered}
\end{equation}

The second detuning component is nonlinearly proportional to the deflection amplitudes due to the Duffing or rather Kerr nonlinearity. In our system the geometric nonlinearities and $F_\text{OMG}$ both contribute to this Duffing term. Both components are defined by

\begin{equation}
	\begin{gathered}
	\beta^{\text{i,GNL},\text{RWA}}_\text{1,1,1,1} := \frac{3\hbar (\beta_{\text{1,1,1,1}}^\text{i,GNL}) }{4 (\omega_\text{i,1}^\text{2} m_\text{i,1}^\text{2})},
	\end{gathered}
	\label{eqn:detuningBetaGNL}
\end{equation}

\begin{equation}
	\begin{gathered}
	\beta^{\text{i,OMG},\text{RWA}}_\text{1,1,1,1} :=  \frac{-3\hbar( P_\text{dc} L g_\text{3}) }{4 (\omega_\text{i,1}^\text{2} m_\text{i,1}^\text{2})}.
	\end{gathered}
	\label{eqn:detuningBetaOMG}
\end{equation}

Both are eventually summed to yield the Duffing or rather Kerr nonlinearity contribution to the detuning according to

\begin{equation}
	\begin{gathered}
	\delta \beta^\text{i} := (\beta^\text{i,GNL,RWA}_\text{1,1,1,1} + \beta^\text{i,OMG,RWA}_\text{1,1,1,1}) |a_\text{i}|^\text{2}.
	\end{gathered}
\end{equation}

To avoid confusion with the non-scaled $\beta_\text{1,1,1,1}^\text{i,GNL}$, we add the superscript (RWA) to denote that they are the coefficients after rescaling according to eq.~(\ref{eqn:detuningBetaGNL}) and eq.~(\ref{eqn:detuningBetaOMG}).

The third component contributing to the detuning is the Cross-Kerr nonlinearity which also nonlinearly detunes the WG oscillation frequency as a function of the NS oscillation deflection and vice-versa. However, unlike $\delta \beta^\text{i}$ it is solely promoted by $F_\text{OMG}$ and does not have a geometric nonlinearity component. We formulate the Cross-Kerr contribution to the detuning as

\begin{equation}
	V^\text{OMG} := \frac{-3 \hbar P_\text{dc} L g_\text{3}}{2 \omega_\text{WG,1} \omega_\text{NS,1} m_\text{WG,1} m_\text{NS,1}},
	\label{eqn:ROWCrossKerr}
\end{equation}

and rescale it for for the WG and the NS further in

\begin{equation}
	\begin{gathered}
	\delta V_\text{WG} :=  V^\text{OMG}|a_\text{NS}|^2,
	\end{gathered}
\end{equation}

\begin{equation}
	\begin{gathered}
	\delta V_\text{NS} :=  V^\text{OMG}|a_\text{WG}|^2.
	\end{gathered}
\end{equation}

To complete the definitions used in eqs.~(\ref{eqn:AWGreal})-(\ref{eqn:ANScomplex}) we finally define the damping of the beams by 

\begin{equation}
d^\text{RWA}_\text{i} = \frac{\omega_\text{i,1}}{2 * Q_\text{i}}.
\label{eqn:ROWDampingTerm}
\end{equation}

Overall, eqs.~(\ref{eqn:AWGreal})-(\ref{eqn:ANScomplex}) reproduce the equations of the well-known degenerate optical parametric oscillator (DOPO). The steady-state phase diagram of the DOPO has been extensively studied \cite{walls2008quantum,carmichael2009statistical}. Most importantly, we differentiate between the trivial phase with $a_\text{NS}=0$ and the non-trivial phase with $a_\text{NS} \neq 0$. For parameter cases satisfying $\delta_\text{WG} \delta_\text{NS} > d^\text{RWA}_\text{WG} d^\text{RWA}_\text{NS}$ the steady state phase diagram shows bistability of the two phases. Moreover, for parameter cases satisfying $\delta_\text{WG} \delta_\text{NS} < -d^\text{RWA}_\text{WG} d^\text{RWA}_\text{NS} -({d^\text{RWA}_\text{WG}}^2 + \delta_\text{WG}^2)/2$ the phase diagram exhibits regions with amplitude modulated stationary states or rather limit-cycles. In order to study such a rich phase diagram on an actual physical device, it is essential to reach the so-called critical point which separates the trivial and the non-trivial phases within the operational ranges of the device. Therefore, we calculate the critical steady state oscillation amplitude of the WG by setting the time-derivatives in eqs.~(\ref{eqn:AWGreal})-(\ref{eqn:ANScomplex}) to zero. At this point it makes most sense to state the critical amplitude in terms of the actual oscillatory deflection $A_\text{WG}$ at angular frequency $\omega_\text{d}$ defined via $q_\text{WG} = A_\text{WG} \sin(\omega_\text{d} t)$.  It is given by  

\begin{equation}
A^\text{cr}_\text{WG} = 2 \frac{\omega_\text{NS} m_\text{NS}}{g_\text{2}  P_\text{dc} L} (\sqrt{\delta_\text{NS}^\text{2} + d^\text{RWA}_\text{NS}}).
\label{eqn:CAmp}
\end{equation}

Further, it is possible to transform the critical amplitude into critical mechanical acceleration $a^\text{cr}_\text{vib}$ and modulated optical power amplitudes $P^\text{cr}_\text{ac}$ - which are considered as external excitation sources in the equations of motion - using the following relations 

\begin{equation}
a^\text{cr}_\text{vib} = \frac{K_\text{WG,1}}{0.52 \rho A_\text{WG} L Q_\text{WG}} A^\text{cr}_\text{WG},
\end{equation}

\begin{equation}
P^\text{cr}_\text{ac} = \frac{K_\text{WG,1}}{g_\text{0} L Q_\text{WG}} A^\text{cr}_\text{WG}.
\end{equation}

\subsubsection{Dynamic analysis results\label{sec:Dynamics}}

For all the results presented in this subsection, we use a constant biasing optical power $P_\text{dc} = 3~\text{mW}$ to set the system to an operating point where a non-linear optical response is expected, yet, not too close to the collapse point, as indicated by the gray line in Fig.~\ref{fig:NSNSFF}. This is essential to have a measurable nonlinear optical phase or transmission responses modulated by the optical input power or the external vibratory force.

\begin{figure}
	\centering
	\includegraphics[width=1.0\columnwidth]{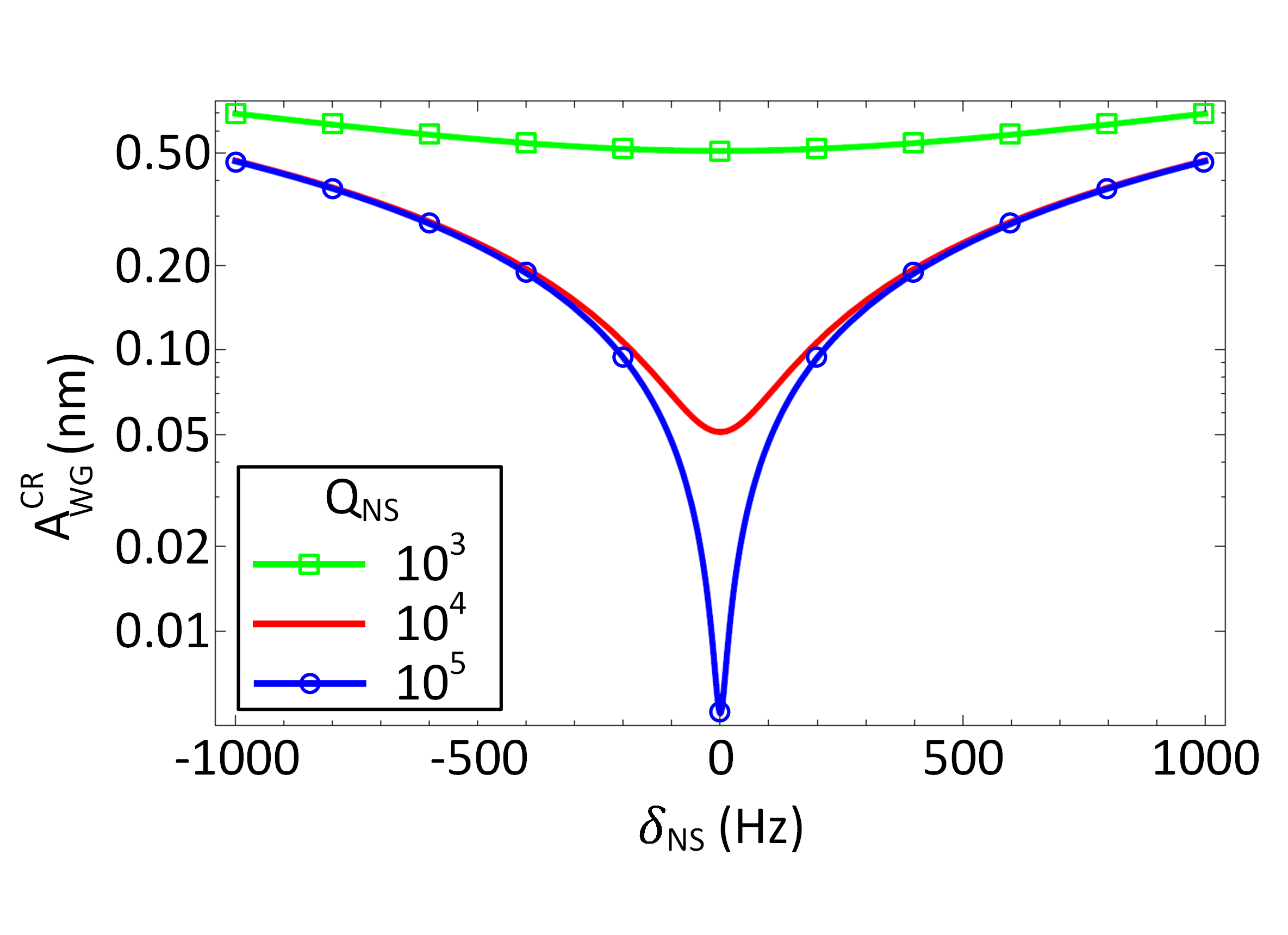}
	\caption{Critical waveguide oscillation amplitude as a function of the NS detuning $\delta_\text{NS}$ with $P_\text{dc} = 3$\,mW for different quality factors $Q_\text{NS}$.   \label{fig:CPC_FOpt}}
\end{figure}

The fundamental mode frequencies $\omega_{\text{WG},1}$, and $\omega_{\text{NS},1}$ are calculated from Euler-Bernoulli's equation to be of 346 KHz and 173 KHz, respectively. For dynamic excitation, the excitation pump frequency is assumed to match the waveguide fundamental mode frequency such that $\omega_{d}=\omega_{\text{WG},1}$. Figure~\ref{fig:CPC_FOpt} shows the critical oscillation amplitude of the waveguide against the detuning $\delta_\text{NS}$ following eq.~(\ref{eqn:CAmp}) at different quality factors. At zero detuning the critical amplitude drops far below $1$\,nm making it easier to practically reach the SPDC regime. However, even away from perfect resonance, i.e. $\delta_\text{NS}=0$, the critical amplitude remains below $1$\,nm, thus well within reach of the stable operation regime as well as within the validity of our approximative model.

\begin{figure}
	\centering
	\includegraphics[width=1.0\columnwidth]{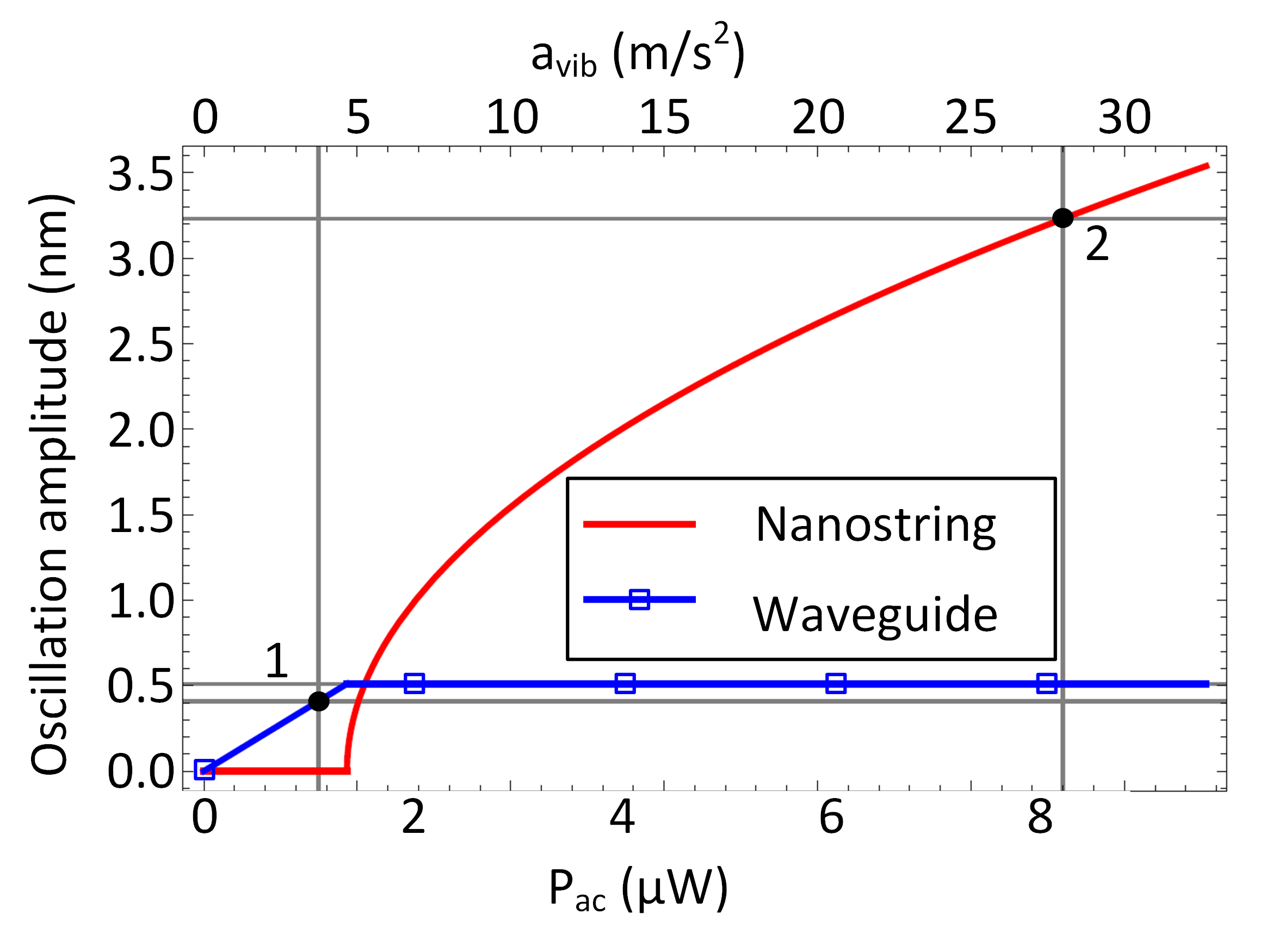}
	\caption{Steady state phase diagram of the nanostring and the waveguide for $Q_\text{NS}=10^3$. The abscissa shows the two proposed dynamic excitation conditions: Top axis) Mechanical shaker acceleration. 2) Lower axis) Optical AC power. The marked points 1, and 2 represent the parameters used further to plot optical phaseshift in Fig.~\ref{fig:PhaseDiagram}. \label{fig:SSAmps}}
\end{figure}

Next, we use eqs.~(\ref{eqn:AWGreal})-(\ref{eqn:ANScomplex}) to determine the steady state amplitudes for the perfectly mode matched case, i.e. $\delta_\text{WG}=\delta_\text{NS}=0$. Figure~\ref{fig:SSAmps} shows the results for the WG and the NS as a function of either the external vibratory excitation $a_\text{vib}$ or the modulated light power $P_\text{ac}$.

The dissipative phase transition from the trivial to the nontrivial solution occurs at the bifurcation point which scales as $1/\alpha$. For the given geometry, $\alpha/d_\text{WG} \approx \frac{2}{3}10^{-7} Q_\text{WG} P_\text{dc} ~1/\,$mW. For an static input optical power $P_\text{dc} = 3 ~\text{mW}$ and a modulated optical power $P_\text{ac} = 1.2\,\mu$W, and for mechanical WG and NS quality factors of $10^\text{3}$, the bifurcation point is reached at an acceleration of $5$\,m$/$sec$^2$. This corresponds to a critical amplitude only of $0.5$\,nm. 

\begin{figure}
	\includegraphics[width=\columnwidth]{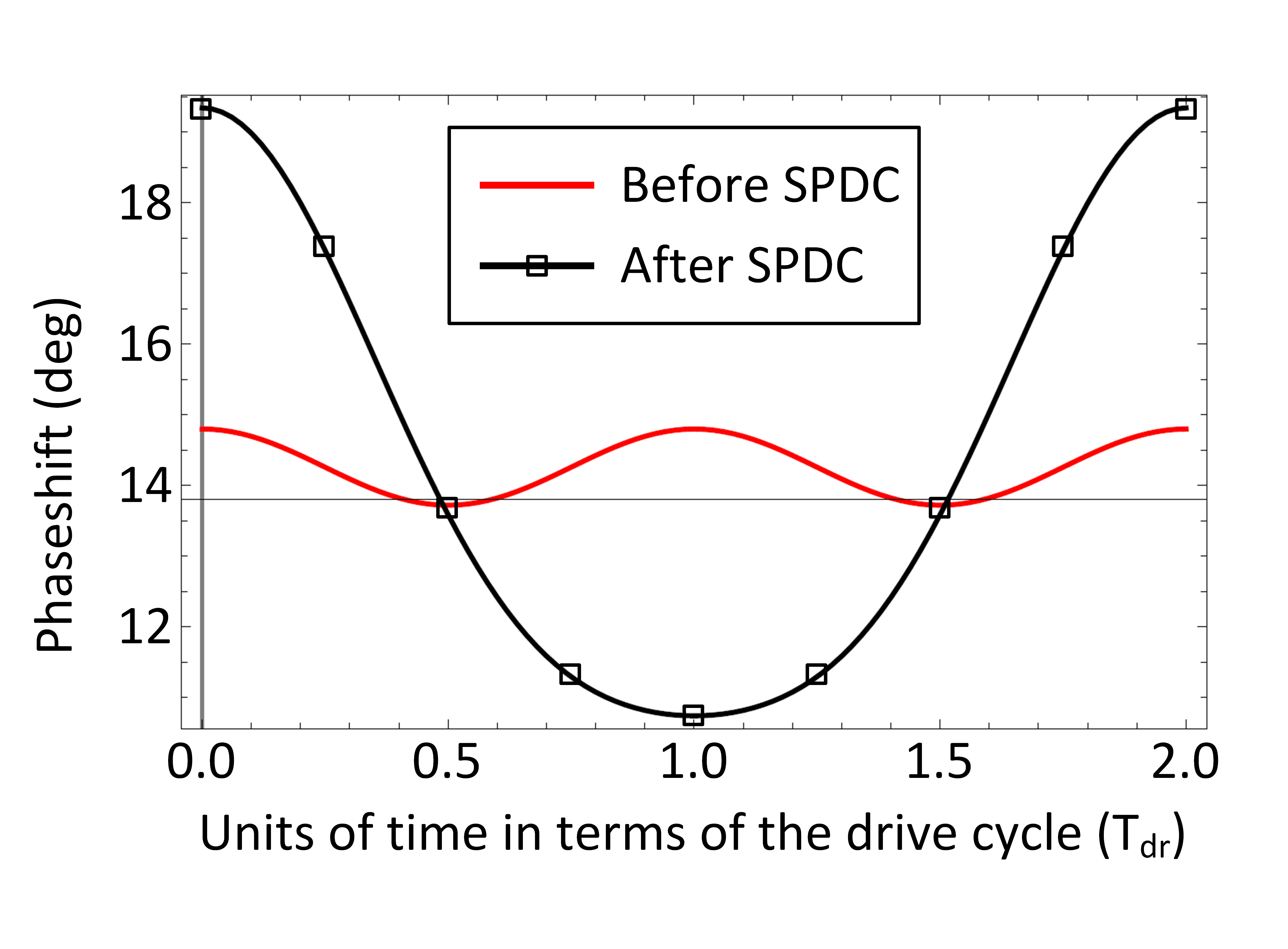}
	\caption{Optical phaseshift under oscillatory excitation. Red: Before SPDC, the oscillation is dominated by $\omega_\text{d}$, corresponding to point (1) on the phase diagram (Fig.~\ref{fig:SSAmps}). Black: After SPDC, the oscillation is dominated by $\omega_\text{d}/2$ and the full wave component $\omega_\text{d}$ is observed as an envelope at the peaks and troughs of the oscillation, corresponding to point (2) on the steady state phase diagram (Fig.~\ref{fig:SSAmps}). \label{fig:PhaseDiagram}}
\end{figure}

Finally we plot the time-transient response of the optical phaseshift produced form the waveguide system under dynamic excitation in Fig.~\ref{fig:PhaseDiagram}. Before the phase transition the output optical phase oscillation is dominated by $\omega_\text{d}$, or the full wave component. However after the phase transition, the NS dominates the optical phaseshift response and we see that not only the amplitude of the phase response increases but also the frequency drops to the half wave component $\omega_\text{d}/2$. For a sufficiently large observable difference between the phase response after the onset of the SPDC, and before, we plot the case after SPDC with an oscillation amplitude of $3$\,nm occurring at $P_\text{ac} = 8\,~\mu$W or mechanical external vibration with an acceleration of $28$\,m$/$s$^2$, when $Q_\text{NS}= 10^3$. In Fig.~\ref{fig:PhaseDiagram} the x-axis is in terms of the of the excitation period $T_\text{dr} = 2\pi/\omega_\text{d}$.

Due to fabrication imperfections it is very difficult to reach the perfectly mode matched situation ($\delta_\text{WG}=\delta_\text{NS}=0$) in an experiment just from the design conception of the beams. In addition, the effective resonance frequencies will also be influenced by the deflection amplitudes and laser powers. These effects have to be considered in order to have full experimental control over the detuning parameters which ultimately determine the dissipative phases of our system. Therefore,
we list the values of the different contributions to the detuning parameters in Table~\ref{tab:Detuning}. The largest contributing components to the detunings are
the linear frequency shifts originating from the optomechanical gradient force. Remarkably, we find that with the optical bias power $P_\text{dc}$ in the mW range, we can shift the frequency on the order of kHz. In contrast, the deflection-dependent nonlinear frequency shifts (Duffing, Kerr, and cross-Kerr) are negligible within our parameter ranges, thus the drive frequencies do not necessarily need to be adjusted during a ramp-up of the oscillation amplitudes. While, the WG detuning $\delta_\text{WG}$ can be adjusted by the external drive frequency, the NS detuning is hard to control simultaneously. 
However, one can imagine to use an electrode beneath the NS
to electrostatically control the resonance frequency of the NS~\cite{bib:Unterreithmeier2009}.

\begin{table}
	\caption{\label{tab:Detuning}Detuning parameters.}
	\begin{ruledtabular}
		\begin{tabular}{lll}
			Detuning component & WG    & NS     \\ \hline
			\begin{tabular}[c]{@{}l@{}}Linear frequency shift from $F_\text{OMG}$\\ (Hz/mW)\end{tabular} & -722  & -2891  \\ 
			\begin{tabular}[c]{@{}l@{}}Duffing shift from to GNL\\ (Hz/nm$^2$)\end{tabular}                                    & 0.46  & 0.92   \\
			\begin{tabular}[c]{@{}l@{}}Kerr shift from $F_\text{OMG}$\\ (mHz/(nm$^2\cdot$ mW))\end{tabular}                       & -14.5 & -58    \\
			\begin{tabular}[c]{@{}l@{}}Cross-Kerr shift from $F_\text{OMG}$\\ (mHz/(nm$^2\cdot$mW))\end{tabular}                 & -29   & -116.1
		\end{tabular}
	\end{ruledtabular}
\end{table}

\section{Conclusion}

We theoretically explored a waveguide system consisting of two suspended, parallel nanobeams, a light-carrying waveguide and a nanostring of half the waveguide's width. As a result of the attractive optomechanical gradient force exerted by the evanescent field of the waveguide onto the nanostring, the input optical power injected into the waveguide can be employed to control the nonlinear coupling between the fundamental vibrational modes of the two nanobeams. More specifically, we show that a tunable 3-wave coupling between the fundamental modes of two doubly clamped nanophotonic beams arises, which is proportional to the input power.

We find that the model describing the response of the optomechanical waveguide system corresponds to the well-known model of the degenerate parametric oscillator. For the nanobeam width ratio of $1:2$ under investigation, the 1:2 internal resonance between the fundamental modes supports degenerate spontaneous parametric down-conversion (SPDC). For beams with a length of $100\,\mu$m and an initial separating gap $130$\,nm, we show that for mechanical quality factors in the range of $10^3$ the critical point of SPDC lies below $1$\,nm of the wider beam's oscillation amplitude for an input optical power in the range of milliwatts. This is well in reach of state of the art experiments, such that our proposal can be well implemented with currently available experimental techniques \cite{Fong:11,van2012ultracompact,quack2019mems,quack2019tue2}. Its physical realization with mechanical structures allows for a high control of the system parameters, therefore, enabling the study of the full dissipative phase diagram.

\addPDS{From the general point of view, the investigations presented in this paper show that the optomechanical gradient forces in nanophotonic waveguides are sufficiently strong already for reasonable light intensities. In addition to actuation and readout~\cite{cai2012nano}, the exploitation of concepts such as frequency tuning for mode matched operation~\cite{fedder2015resonant,nitzan2015self} or mechanical parametric amplification~\cite{rugar1991mechanical} allow the possibility to design sensors~\cite{westerveld2020sensor} and actuators with different functionalities such as for example gyroscopes or accelerometers based on all optical or maybe even hybrid electrical/optical working principles. Overall, the high controllability via the input laser powers and the in-situ tuneability of the optical-gap dependent nonlinearities offer large design flexibilities which are comparable with traditional NEMS and MEMS design but with possible advantages in performance, cost, and reliability especially when aiming for designs on the nano-scale.}

\begin{acknowledgments}
	Mohamed Ashour acknowledges support from the European Union’s Horizon 2020 Programme for Research and Innovation under grant agreement No. 722923 (Marie Curie ETN - OMT).
\end{acknowledgments}

\appendix

\section{Optical-Read out equations \label{App:OpticalReadout}}

We start by representing the output field amplitude at the side of the MZI without the system of the WG and the NS by 

\begin{equation}
E_\text{1} = \frac{E_\text{0}}{2} e^{-i \frac{2 \pi L}{ \lambda} n_\text{eff0,1}},
\label{eqn:Theory:MZI_1}										
\end{equation}

On the side with the WG and the NS, a geometry dependent field amplitude would be produced at its output represented by

\begin{equation}
E_\text{2} = \frac{E_\text{0}}{2} e^{-i \frac{2 \pi}{ \lambda} \int_\text{0}^\text{L} n_\text{eff}(z)dz},
\label{eqn:Theory:MZI_2}								
\end{equation} 

assuming a constant input power of $P_\text{in}$. $E_\text{0}$ is the amplitude of the input optical power to the MZI, and $n_\text{eff0,1}$ is the refractive index of the optical waveguide in a silicon slab within silicon dioxide, thus representing the side of the MZI without the WG and the NS. 

We then add a trivial term of $n_\text{eff2,0}L - n_\text{eff2,0}L$ to the exponents in equations \ref{eqn:Theory:MZI_1}, and \ref{eqn:Theory:MZI_2} while summing both to obtain 

\begin{subequations}
	\label{eqn:Theory:MZI_4}
	\begin{align}
	E_\text{3} = \frac{E_\text{0}}{2} e^{-i \frac{2 \pi}{ \lambda}(\int_\text{0}^{L}n_\text{eff0,2}(Z)dz + n_\text{eff0,2}L-n_\text{eff0,2}L)} + \nonumber \\ \frac{E_\text{0}}{2} e^{-i\frac{2\pi}{\lambda}(n_\text{eff0,1}L + n_\text{eff0,2}L-n_\text{eff0,2}L)}. 	\tag{\ref{eqn:Theory:MZI_4}}
	\end{align}					
\end{subequations}

This equation represents the output electric field. $n_\text{eff2,0}$ is the total effective index at the initial gap in the side with the optomechanical system. L is the coupling length between WG and NS, which is also identical to the length of the other side of the MZI.

Now we define the quantity $\delta \phi$ representing the optical phase-shift between both sides of the interferometer as described earlier by eq.~(\ref{eqn:Theory:MZI_5}). The other quantity that is relevant here is the initial phaseshift acquired by a propagating mode through the system of the WG and the NS which is formulated using 

\begin{equation}
\phi_\text{0} = \frac{2\pi}{\lambda}n_\text{eff0,2}.
\label{eqn:Theory:MZI_17}
\end{equation}

The notation $\delta_\text{imbalance}$ denotes the phaseshift imbalance between both sides of the MZI at the initial gap and is defined as

\begin{equation}
\delta_\text{imbalance} = \frac{2\pi}{\lambda}(n_\text{eff0,2} - n_\text{eff0,1})L.
\label{eqn:Theory:MZI_18}
\end{equation}

We then obtain the total electric field at the output port of the MZI using 

\begin{equation}
E_\text{3} = \frac{E_\text{0}}{2} e^{-i (\Delta \Phi(P) + \phi_\text{0})} + \frac{E_\text{0}}{2}e^{-i( \phi_\text{0} + \delta_\text{imbalance})}.
\label{eqn:Theory:MZI_14}
\end{equation}

By substitution of eqs.~(\ref{eqn:Theory:MZI_5}), (\ref{eqn:Theory:MZI_17}), and \ref{eqn:Theory:MZI_18} into eq.~(\ref{eqn:Theory:MZI_4}), we simplify further to arrive at the form 

\begin{equation}
E_\text{3} = \frac{E_\text{0}}{2} e^{-i (\phi_\text{0} + \delta_\text{imbalance})}(1 + e^\text{-i ($\Delta \phi - \delta_\text{imbalance}$)}).
\label{eqn:Theory:MZI_6}
\end{equation}

Next we obtain the output transmission at the output port using

\begin{subequations}
	\label{eqn:Theory:MZI_12}		
	\begin{align}
	T = \big{|}\frac{E_\text{3}}{E_\text{0}}|^\text{2} = \frac{1}{4} |e^{-i (\phi_\text{0} + \delta_\text{imbalance})}\big|^2 \nonumber \\ \big|1 + e^\text{-i ($\Delta \phi - \delta_\text{imbalance}$)}\big|^2,
	\tag{\ref{eqn:Theory:MZI_12}}
	\end{align}
\end{subequations} 

which is further simplified to 

\begin{subequations}
	\label{eqn:Theory:MZI_7}	
	\begin{align}
	T(P_\text{in})= \big|\frac{E_\text{3}}{E_\text{0}}\big|^{2} = \frac{1}{4}\big|1+cos(\Delta \phi(P_\text{in}) - \delta_\text{imbalance}) + \nonumber \\ isin(\Delta \phi(P_\text{in}) - \delta_\text{imbalance})\big|^{2}.
	\tag{\ref{eqn:Theory:MZI_7}}
	\end{align}		
\end{subequations}

Finally the transmission reduces to

\begin{equation}
T(P_\text{in}) = \big|\frac{E_\text{3}}{E_\text{0}}\big|^2 = \frac{1}{2}(1+cos(\Delta \phi(P_\text{in}) - \delta_\text{imbalance})).
\label{eqn:Theory:MZI_10}
\end{equation}

We then multiply the transmission by the input power to get the following nonlinear response of the system 

\begin{equation}
P_\text{out} = P_\text{in} \frac{1}{2}(1+cos(\Delta \phi(P_\text{in}) - \delta_\text{imbalance})).
\label{eqn:Theory:MZI_11}
\end{equation}

Note that for the analysis we included the imbalance term stemming from the different refractive indices at both sides of the MZI. However, it results in merely offsetting the position of the
full $\pi$ phaseshift, thus we can safely neglect it in our analysis presented in the main text of the paper.

\section{Mechanical parameters for modal representation of Euler-Bernoulli's beam equation \label{App:EBP}}

\begin{figure}[h!]
	\centering
	\includegraphics[width=1.0\columnwidth]{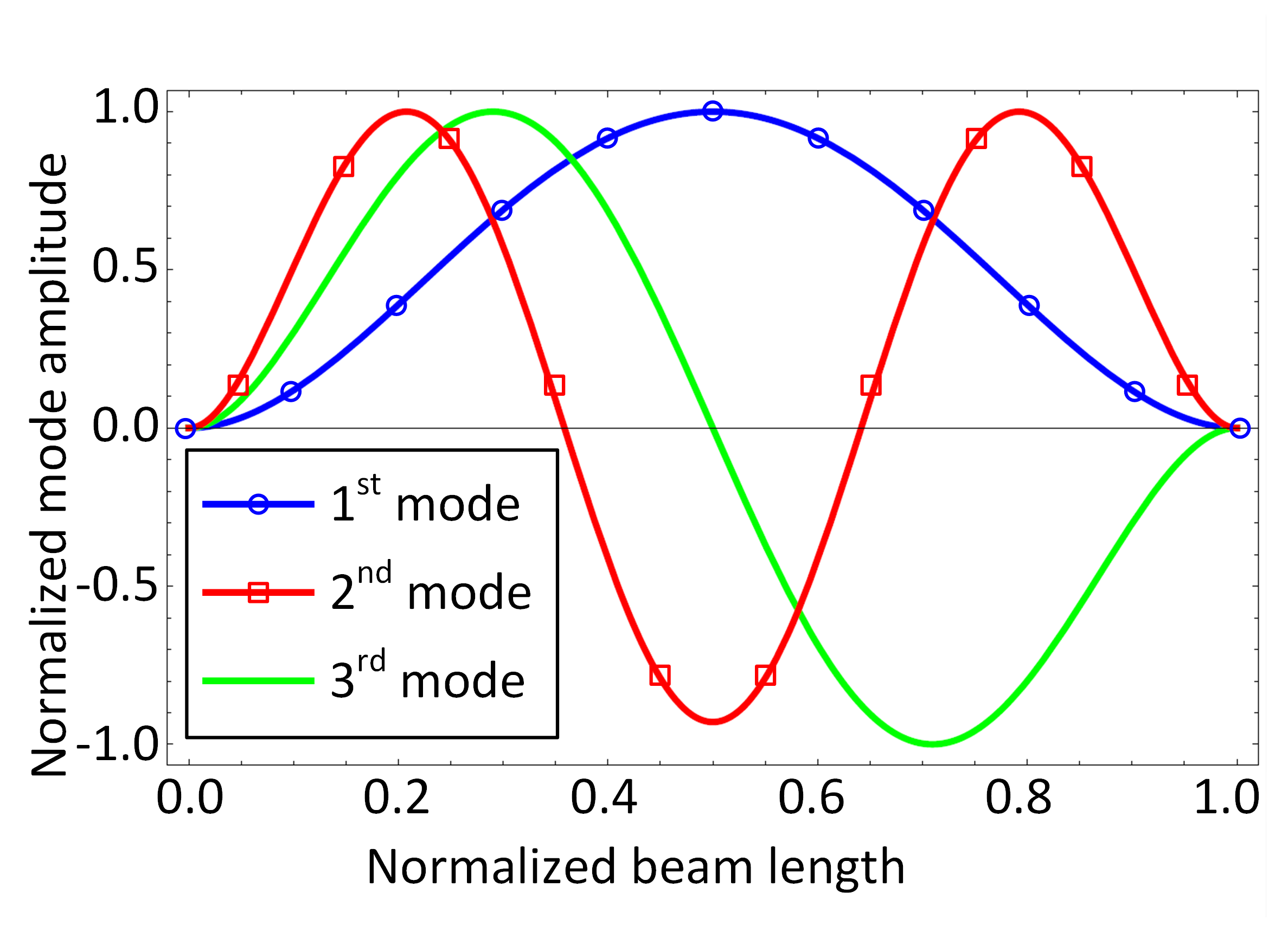}
	\caption{Mode shapes of the doubly-clamped Euler-Bernoulli beam. \label{fig:Modes}}
\end{figure}

The modal shape functions of the doubly-clamped Euler-Bernoulli beam are given by \begin{subequations}
	\label{eqn:modeshapes}
	\begin{align}
		S_\text{n}(z) = N_\text{n}(\cosh(r_\text{n}z)-\cos(r_\text{n}z)- \nonumber \\ \frac{\cosh(r_\text{n})-\cos(r_\text{n})}{\sinh(r_\text{n})-\sin(r_\text{n})}(\sinh(r_\text{n}z)-\sin(r_\text{n}z)),
		\tag{\ref{eqn:modeshapes}}
	\end{align}
\end{subequations}

with the normalization factors $N_\text{n}$ such that the maximal deflection of $S_\text{n}$ equals unity. We illustrate the first three shape functions in Fig.~\ref{fig:Modes}.  

In order to represent the equations of motions in the modal form for the n-th modal amplitude $q_\text{i,n}$, we multiply the field form of the equation of motions shown in eq. (\ref{eq:EBTD}) for the WG and the NS by the linear normal mode shape $S_\text{n}(z)$ and then integrate the equations according to $\int_\text{0}^\text{1}Ldz$. We can then define the modal mass $m_\text{i,n}$, damping $d_\text{i,n}$, stiffness $k_\text{i,n}$, Duffing $\beta^{i}_\text{n,m,l,k}$, and $F_\text{i,OMG}$ using the following set of equations

\begin{subequations}
	\label{eq:ModalMass}
	\begin{align}
		m_\text{i,n} := \rho A_\text{i}L \int_\text{0}^\text{1} S^\text{2}_\text{n}(z) dz, \tag{\ref{eq:ModalMass}}
	\end{align}
\end{subequations}

\begin{subequations}
	\label{eq:ModalDamping}	
	\begin{align}
		d_\text{i,n} := dA_\text{i}L \int_\text{0}^\text{1} S_\text{n}^\text{2}(z) dz,		\tag{\ref{eq:ModalDamping}}
	\end{align}
\end{subequations}

\begin{subequations}
	\label{eq:ModalStiffness}		
	\begin{align}
		K_\text{i,n} := \frac{YA_\text{i}h^\text{2}}{12L^\text{3}}
		{\sum}_\text{m=1}^\text{$\infty$}{\int}_\text{0}^\text{1}S_\text{n}(z)S_\text{n}^{''''}(z)dz,
		 \tag{\ref{eq:ModalStiffness}}		
	\end{align}
\end{subequations}

\begin{subequations}
	\label{eq:ModalDuffing}			
	\begin{align}
		{\beta}^\text{i,GNL}_\text{n,m,l,k}:=\frac{YA_\text{i}}{2L^\text{3}} \int_\text{0}^\text{1}S^{'}_\text{n}(z) S^{'}_\text{m}(z) dz \int_\text{0}^\text{1}S^{'}_\text{l}(z) S^{'}_\text{k}(z) dz,
		\tag{\ref{eq:ModalDuffing}}				
	\end{align}
\end{subequations}

\begin{subequations}
	\label{eq:ModalFOMG}				
	\begin{align}
		F^\text{i}_\text{OMG,n}(z,t):= L\int_\text{0}^\text{L}S_\text{n}(z)F^\text{i}_\text{OMG}(G_{0}- \nonumber \\ (u_\text{WG}(z,t)+u_\text{NS}(z,t)))dz.
		\tag{\ref{eq:ModalFOMG}}						
	\end{align}
\end{subequations}

Due to the orthogonality, the modes completely decouple in the linear regime. The geometrical symmetry as well of the beams makes 3-wave forces $\alpha_{n,m,l}^{GNL}q_{n}q_{m}q_{l}$  originating from geometric nonlinearity negligible. The 3-wave coupling term is however promoted by the optomechanical gradient force and is calculated in section~\ref{subsec:ROW}.
\\
\section{Finite element analysis \label{App:FEM}}

To verify the effectiveness of the modal approximation, we implement a 2D Finite Element Model (FEM) to simulate the static deflection of the beams in the system of the WG and the NS under static optical pumping. We compare our static deflection obtained from the simulation to the modal approach.

To address the need for the input power-dependent gap between the WG and the NS each two opposite mesh points in boundaries ($\Gamma_1$, $\Gamma_2$) in (Fig.~\ref{fig:SlotWaveguide}) are treated as dependent variables rather than independent inputs. As a highly nonlinear problem arises from this setting, a single iteration will only converge if the restoration forces across the waveguides have completely balanced the optical force in a specific iteration. The solution will be stable as long as this condition is achieved. Once a case occurs where the restoration forces are unable to balance the increasing optical force, the beams would collapse and the minimum gap between the two waveguides would be higher than the initial gap between them.

We used COMSOL Multiphysics as it provides boundary coupling and the desired solver parameters. The maximum number of iterations per power value was set to 500 to avoid pre-mature divergence, while allowing enough iterations to reach a stable solution. The numerical damping factor of the values used in the nonlinear solver from iteration to iteration was set to $10^{-9}$.

\bibliography{bib}
\end{document}